\documentclass[amsmath,amssymb, aps, bibtex, prl,letter,reprint,
 shortbibliography
]{revtex4-2}
\usepackage{verbatim}
\usepackage{hyperref} 
\usepackage{svg}
\usepackage{graphicx}
\usepackage{dcolumn}
\usepackage{bm}
\usepackage{amsmath}
\usepackage{enumerate}
\usepackage[normalem]{ulem}
\usepackage{float}
\usepackage[english]{babel}

\newcommand{\be}{\begin{equation}}
\newcommand{\ee}{\end{equation}}
\newcommand{\bae}{\begin{eqnarray}}
\newcommand{\eae}{\end{eqnarray}}

\begin{document}

\preprint{APS/123-QED}


\title{Frequency-dependent covariance reveals critical spatio-temporal patterns of synchronized activity in the human brain.}
\author{Rub\'en Calvo} 
\affiliation{Departamento de Electromagnetismo y F{\'\i}sica de la Materia and Instituto Carlos I
de F{\'\i}sica Te{\'o}rica y Computacional. Universidad de Granada.E-18071, Granada, Spain}
 \author{Carles Martorell}
\affiliation{Departamento de Electromagnetismo y F{\'\i}sica de la Materia and Instituto Carlos I
de F{\'\i}sica Te{\'o}rica y Computacional. Universidad de Granada.E-18071, Granada, Spain}
\author{Guillermo B. Morales}
\affiliation{Departamento de Electromagnetismo y F{\'\i}sica de la Materia and Instituto Carlos I
de F{\'\i}sica Te{\'o}rica y Computacional. Universidad de Granada.E-18071, Granada, Spain}
\author{Serena Di Santo}
\author{Miguel A. Mu\~noz} 
\affiliation{Departamento de Electromagnetismo y F{\'\i}sica de la Materia and Instituto Carlos I
de F{\'\i}sica Te{\'o}rica y Computacional. Universidad de Granada.E-18071, Granada, Spain}

\date{\today}

\begin{abstract}
Recent analyses combining advanced theoretical techniques and high-quality data from thousands of simultaneously recorded neurons provide strong support for the hypothesis that neural dynamics operate near the edge of instability across regions in the brain. However, these analyses, as well as related studies, often fail to capture the intricate temporal structure of brain activity as they primarily rely on time-integrated measurements across neurons. In this study, we present a novel framework designed to explore signatures of criticality across diverse frequency bands and construct a much more comprehensive description of brain activity. Additionally, we introduce a method for projecting brain activity onto a basis of spatio-temporal patterns, facilitating time-dependent dimensionality reduction. Applying this framework to a magnetoencephalography dataset, we observe significant differences in both criticality signatures and spatio-temporal activity patterns between healthy subjects and individuals with Parkinson's disease.
\end{abstract}

\maketitle

Nearly a century after H. Berger first recorded endogenous brain activity, deciphering its origin and functional significance remains a crucial challenge. Technological advances enabling the simultaneous recording of the activity of thousands of individual neurons allowed to shift the focus from the study of single neurons to large neuronal ensembles and their collective dynamical behavior, underscoring the need for developing a Statistical Mechanics framework for brain networks \cite{Bassett}. A widely discussed conjecture in this context posits that mathematical models of brain activity need to operate near a critical point or ``close to the edge of instability" to generate activity patterns akin to those observed experimentally, both at microscopic (e.g., electrophysiology or calcium imaging) and macroscopic (functional magnetic resonance imaging (fMRI) or magneto/electro-encephalography (M/EEG)) scales \cite{Cabral, Haimovici, RMP, Byrne, Plenz-review}. Crucially, critical behavior has been linked to functional capabilities related to the processing, storage, and transmission of information \cite{Shew2013, Shew2015, Byrne, Plenz-review} and deviations from criticality have been associated with pathology \cite{Massobrio}.
  
Recently, Hu and Sompolinsky (HS) achieved a significant breakthrough in this regard, developing tools to analytically infer the actual dynamical regime from empirical data. In particular, they computed the spectrum of the (long-time integrated) covariance matrix in a model of randomly coupled neurons as a function of its dynamical regime \cite{HS}. Note that such a spectrum is the starting point for dimensionality-reduction techniques like principal component analysis (PCA) \cite{Shlens,Bradde}. Their crucial observation is that the spectrum develops a power-law tail as the overall coupling strength is increased and the system approaches the edge of instability. In contrast, ``sub-critical" dynamics result in a narrower range of eigenvalues and a cutoff in the distribution. These results are robust to modeling details --holding, e.g.,  for non-linear and spiking neuron models--- and sub-sampling effects \cite{HS}. Consequently, it becomes possible to estimate the proximity to the edge of instability from empirical time series by fitting the measured covariance spectrum to the theoretical one \cite{HS}. This strategy has been successfully applied to data from the simultaneous recording of thousands of neurons in different regions of the mouse brain \cite{Steinmetz}, showing that all recorded areas are, to varying degrees, close to the edge of instability \cite{Morales2023}. 
The previous analyses are limited, however, in that they only examine time-integrated covariances, i.e.,
\be   \label{S0}
 \int_{-\infty}^{\infty} d\tau C_{ij}(\tau) \equiv \int_{-\infty}^{\infty} d\tau \int_{-\infty}^{\infty} dt ~ \langle x_i(t) x_j(t+\tau) \rangle 
\ee
where $i,j=1...N$ label different neurons or recording channels, $x_i(t)$ are their firing rates at time $t$, and $\langle \cdot \rangle$ stands for average over independent measurements. In other words, the temporal structure is averaged out. However, whole-brain activity exhibits time-dependent complex oscillations that occur at a broad range of frequency bands ($\alpha, \beta, \gamma,...$) \cite{Buzsaki2006, Buzsaki2023} and coexists with a background of non-periodic ($1/f$) noisy activity \cite{Vojtek2020}.
 Such complex collective rhythms are characterized by transient neuronal synchronization and are believed to be essential for information transmission and integration across regions \cite{Plenz-gamma, Deco-synchro, diSanto, Zhou2020}. Moreover, recent work has analyzed how effective neural communication protocols crucially depend on the distance of the intrinsic dynamics to the edge of oscillatory synchrony \cite{Battaglia2017}.
This underscores the need to extend the mentioned time-integrated analyses \cite{HS} to investigate multi-frequency band aspects of criticality. We tackle this problem by analyzing frequency-dependent covariance matrices, whose elements can be written using the Wiener-Kinchin theorem \cite{Gardiner} as a Fourier transform of $C_{ij}(\tau)$ (see SI):
\begin{equation}
\label{Sw0}
 S_{ij}(\omega)=\int_{-\infty}^{\infty} d\tau C_{ij}(\tau) e^{-i\omega\tau},
\end{equation}
so that $S_{ij}(\omega=0)$ coincides with Eq.(\ref{S0}). The resulting frequency-dependent covariance (FDC) matrix, $\boldsymbol{S}(\omega)$, has been employed to analyze EEG \cite{ji2018data}, MEG \cite{Stam2012} and fMRI \cite{Bullmore2005} data. Here, we go beyond such analyses and examine whether an approach akin to that in \cite{HS,Morales2023} can be expanded to deal with FDC matrices, with the idea of evaluating the distance to a critical regime across frequency bands while generating a basis to describe spatio-temporal propagation patterns in healthy and pathological subjects.

Principal component analysis (PCA) is the best-known dimensionality reduction method for the analysis of multidimensional data \cite{Shlens}. The leading eigenvalues of the covariance matrix describe the directions of maximal variability in the data, allowing for effective low-dimensional descriptions of complex data by projecting onto a subset of leading eigenvectors.
For instance, PCA analyses of fMRI data in resting conditions enabled the inference of a small set of co-activation patterns, called \emph{resting state networks}, offering insights into the default functional connectivity in the brain \cite{RSN2, RSN4}. Likewise, investigations of time-dependent activity under task performance have also resorted successfully to PCA analyses \cite{ames2019motor}.
However, PCA analyses on correlated time series can induce artifacts such as ``phantom oscillations" \cite{shinn2023phantom}. 
Similarly, variants of PCA, such as time-lagged PCA, involve non-Hermitian covariance matrices, resulting in complex eigenvalues and interpretation challenges. Ultimately, a robust tool to investigate multidimensional noisy time-structured data is not yet available. Here, we fill this gap by introducing frequency-dependent PCA (FD-PCA), which exploits the Hermitian FDC matrix $\boldsymbol{S}(\omega)$.

\begin{figure}[tbh!]
    \centering \includegraphics[width=1 \linewidth]{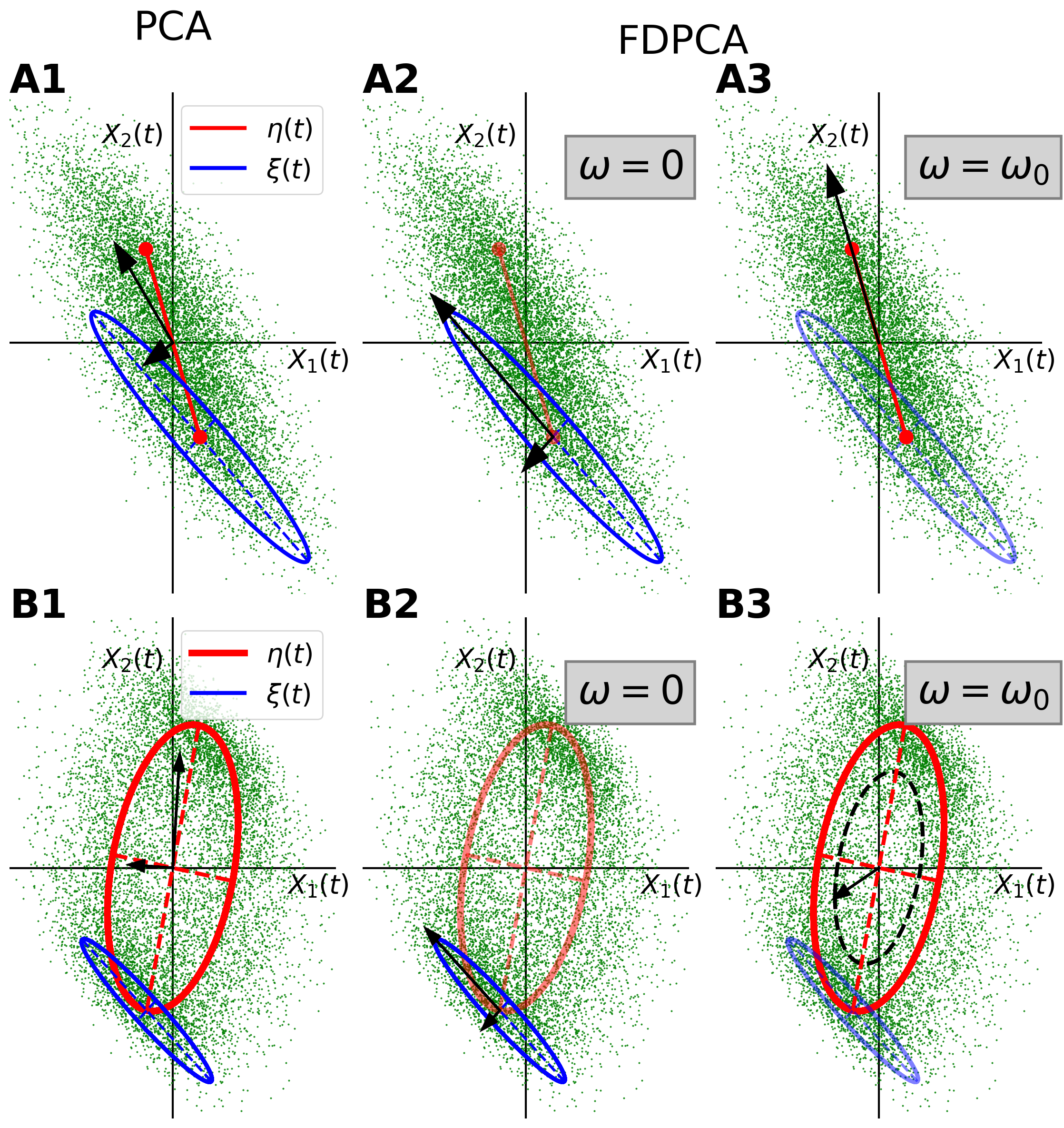}
    \vspace{-0.25cm}
    \caption{\textbf{PCA vs FD-PCA analysis of the toy model described by Eq.(\ref{Example1})}  Upper row: no delay, $\delta=0$, and $\boldsymbol{v}=(-1,3)$. Lower row: delay $\delta=2\pi/1.43$, $\boldsymbol{v}=(-2,4)$. $\rho=-0.8$ in both.
    The blue ellipsoids mark the directions of variability of the noisy component $\boldsymbol{\xi}(t)$, the red curves are the trajectories of the periodic signal, $\boldsymbol{\eta}(t)$, as specified in Eq.(\ref{Example1}), and the black vectors denote the principal directions (PCA in the first column, and FD-PCA in the second and third, at $\omega=0$ and $\omega_0$, respectively). \textbf{A1}/\textbf{B1}: Scatter plot of the dynamical process.  
    The principal direction from PCA mixes the contributions from both noise and signal variability. \textbf{A2}/\textbf{B2}: At $\omega=0$, the periodic signal contribution is filtered out  (fading red color), so that the principal directions of $\boldsymbol{S}(0)$ coincide with those of the noise term. \textbf{A3}/\textbf{B3}: At the characteristic frequency $\omega_0$, noise is filtered out, leaving, in the case without delay (\textbf{A3}), the principal direction of the signal process (only one principal direction is shown). 
    For non-vanishing delays, since the principal eigenvector is complex, its projection onto the real space parametrizes the whole trajectory of the ellipse (black dotted line; the black arrow is the projection at one specific time; see main text). }
\label{fig:PCA-FDPCA}
\end{figure}

 To gain intuition, let us first describe a simple discrete-time dynamical process for two variables, $x_1(t) $ and $x_2(t) $:
\begin{equation}
\label{Example1} 
    \begin{cases}
       x_1(t)  = v_1 \cos(\omega_0 t + \varepsilon) & +   ~~\xi_1(t)  \\
   x_2(t) = v_2 \cos(\omega_0 t + \delta + \varepsilon)& +  ~~\xi_2(t)
    \end{cases}\, ,
\end{equation}
which includes a periodic signal
$\omega_0$, relative delay $\delta$, a random phase $\varepsilon$, and coupling amplitudes denoted by $\boldsymbol{v}=(v_1, v_2)$, along with zero-mean and unit variance Gaussian noise terms \(\xi_i(t)\) (\(i=1, 2\)), which exhibit crossed-correlations $\langle \xi_1(t)\xi_2(t)\rangle=\rho$ and $|\rho|\leq 1$. Thus, the time-lagged correlation matrix $\mathbf{C}(\tau)$ is given by
\begin{equation}
\label{C}
\delta(\tau)\left( \begin{array}{cc}
1 & \rho \\
\rho & 1
\end{array} \right)+
\frac{1}{2}\left( \hspace{-0.20cm}\begin{array}{cc}v_1^2 \cos(\omega_0 \tau) & v_1 v_2 \cos(\omega_0 \tau + \delta) \\
v_1 v_2 \cos(\omega_0 \tau - \delta) & v_2^2 \cos(\omega_0 \tau) 
\end{array} \hspace{-0.15cm}\right),
\end{equation}
needed for standard PCA analyses while, for FD-PCA, using Eq.(\ref{Sw0}) (see SI) 
\be 
\label{Sw}
\boldsymbol{S}(\omega) = \left(  \begin{array}{cc}
    1 & \rho \\
    \rho & 1
\end{array}\right) + 
\frac{\pi \delta(\omega \pm \omega_0)}{2}  
\left(  \begin{array}{cc}
    v_1^2 & v_1 v_2 e^{- i \delta \frac{\omega}{\omega_0}} \\
    v_1 v_2 e^{ i \delta \frac{\omega}{\omega_0}} & v_2^2
\end{array} \hspace{-0.15cm}\right).
\ee
Fig.\ref{fig:PCA-FDPCA} shows scatter plots for one realization of the above process in two different cases: without delay (upper row, $\delta=0$) and with a non-vanishing delay (lower row, $\delta \neq 0$). Blue ellipsoids represent the directions of noise variability (eigenvectors of the first matrix in Eq.\ref{C})), the red curves represent the signal trajectory.
Observe that, as illustrated in Fig.\ref{fig:PCA-FDPCA}A1/B1, a standard equal-time PCA analysis, giving the direction of overall maximal variability (black vectors), mixes up information from signal and noise variance (i.e., the leading eigenvector is a combination of the principal directions of variability from signal and noise). Instead, FD-PCA evaluated at different frequencies can decompose these two contributions. At $\omega=0$, the periodic signal is filtered out, resulting in the principal eigenvectors of \( \boldsymbol{S}(0) \) coinciding with those of the noise covariance (see black arrows in Fig. \ref{fig:PCA-FDPCA}A2/B2). Conversely, at \( \omega =\pm \omega_0 \), the noise contribution is negligible, and \( \boldsymbol{S}(\omega_0) \) becomes a Hermitian complex matrix with eigenvalues \( \lambda_1 =(\pi/2)\delta(0) \Vert \mathbf{v} \Vert^2 \) and \( \lambda_2 = 0 \). Consequently, its associated eigenvectors \( (v_1, v_2 e^{i \delta}) \) and \( (-v_2, v_1 e^{i \delta}) \) are real only if \( \delta = 0 \), in which case the principal direction aligns with the signal vector \( (v_1, v_2) \) (Fig.\ref{fig:PCA-FDPCA}A3). On the other hand, in the presence of delays (\( \delta \neq 0 \)), both eigenvectors become complex 
and fixed up to an arbitrary factor or ``\emph{gauge}", 
 \( e^{i \varphi} \), preserving the modules and eigenvectors.  
Fixing the gauge to $\varphi=\omega_0 t$ ---which is equivalent to Fourier transforming back the mode of frequency $\omega_0$ to the time domain--- the time-dependent projection of the first eigenvector describes an elliptical trajectory, $(v_1 \cos(\omega_0 t), v_2 \cos(\omega_0 t+\delta))$ that precisely reconstructs the input signal (Fig.\ref{fig:PCA-FDPCA}B3). This simple observation is inspiring because it reveals that the complex eigenvectors of the FDC matrix can be exploited to uncover dynamical spatio-temporal patterns from empirical data.

We now shift to a higher-dimensional neural network model with $N$ randomly coupled nodes whose dynamics are described by \cite{Crisanti88, HS}:
\begin{equation}
\label{rate}
    \dot{x}_i(t)=-x_i(t)+ F\left(g \sum_{j=1}^N W_{ij} x_j(t)\right)+\xi_i(t) \ ,
\end{equation}
where $x_i(t)$, is a continuous variable representing the firing rate of the $i$-th neuron, the elements $W_{ij}$ describe the synaptic weights (the matrix $\boldsymbol{W}$ will be specified later), $g$ controls the overall coupling strength, $\xi_i(t)$ is a zero-mean Gaussian white noise term, with $\langle \xi_i(t)\xi_j(s) \rangle=\sigma^2 \delta_{ij}\delta(t-s)$, and $F(x)$ is a gain or response function, which here we assume to be linear (i.e., $F(x)=x$; in SI we show that the results remain robust regardless of this choice).
In this case, the dynamics are stable as long as the eigenvalues $\lambda_{W,i}$ of $\boldsymbol{W}$ are such that the largest one,  $\lambda_{W,max}$, obeys the condition $g \lambda_{W,max} < 1$, so that $\lambda_{W,max}^*=1/g$ sets the threshold for the edge of instability.  The FDC matrix can be easily computed by defining \( \boldsymbol{A} = -(\boldsymbol{I} - g\boldsymbol{W}) \) (where $\boldsymbol{I}$ is the identity matrix) and Fourier transforming Eq.(\ref{rate}), yielding \cite{Pernice,Trousdale,Dahmen-second}:
\begin{equation}
\label{formula}
\boldsymbol{S}(\omega)={(\boldsymbol{A}+i\omega \boldsymbol{I})}^{-1}~ {(\boldsymbol{A}^T-i\omega \boldsymbol{I})}^{-1} \ .
\end{equation}
As a Hermitian matrix, $\boldsymbol{S}(\omega)$ has real eigenvalues and a set of orthogonal, generally complex eigenvectors. Let us now consider two examples with different $\boldsymbol{W}$ matrices.
\begin{figure*}[tbh!]
        \centering
    \includegraphics[width = 1.0\linewidth]{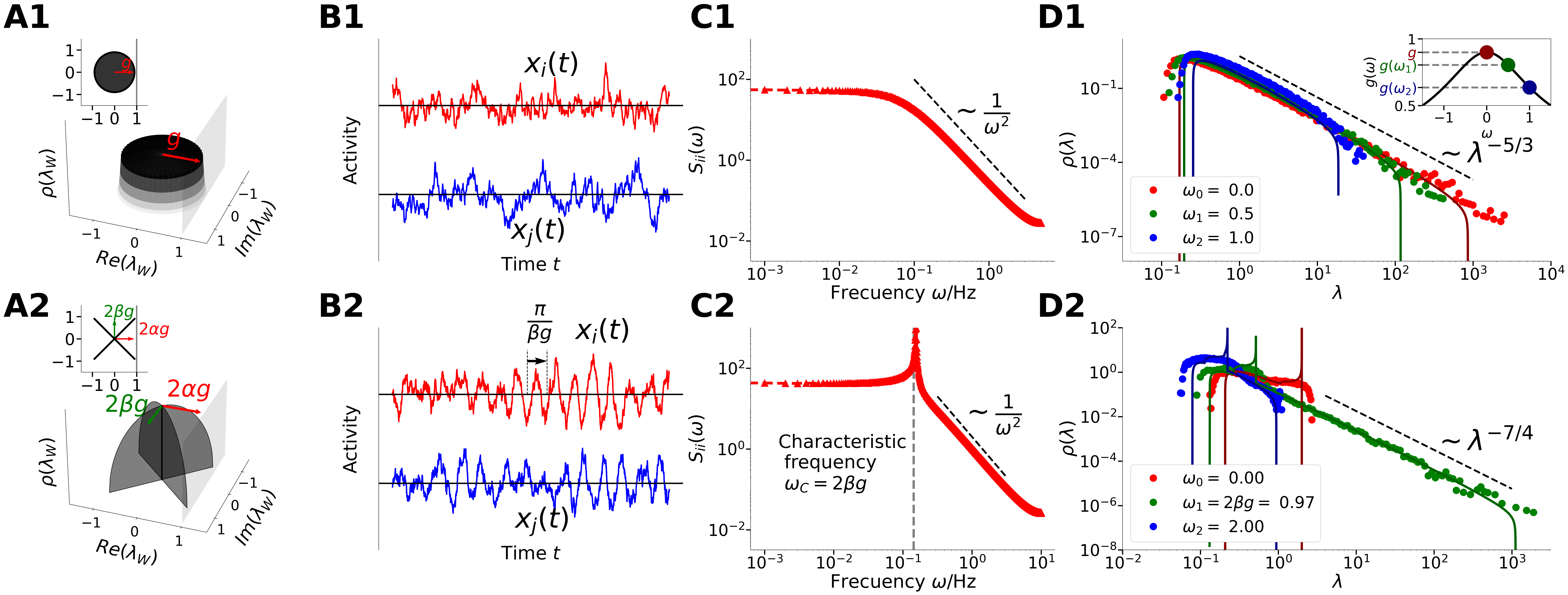}
    \caption{\textbf{Near-critical dynamics at diverse frequency bands for two different 
    connectivity matrices $\boldsymbol{W}$: a random matrix (upper panels) and one with the structure in Eq.(\ref{Oscillatory}).} (lower panels). \textbf{A}. Eigenvalue distribution for a random matrix $g \boldsymbol{W}$, obeying the circle law in the complex plane, with radius $g$ (\textbf{A1})
       and for the matrix defined by Eq.(\ref{Oscillatory}) (\textbf{A2}).  The latter has eigenvalues of the form $(\alpha\pm i\beta)\lambda_J$, where $\lambda_J$ is an eigenvalue of the matrix $J$ --as shown in the inset. The density at each of these lines follows Wigner's semicircle law \cite{vM}. 
 \textbf{B}. Representative time series as obtained from integration of Eq.(\ref{rate}) in the linear case ($N=10^2$ nodes, $10^4$ time units, and integration step $0.1$); observe the presence of noise-sustained quasi-oscillations of characteristic frequency $2\beta g$ in
  \textbf{B2}.
 \textbf{C}. The network-averaged power spectrum exhibits a monotonous decay in the upper case (\textbf{C1}) and a peak at a non-trivial frequency \(2\beta g\) in the lower one (\textbf{C2}).
 \textbf{D}. Eigenvalue distribution of $\boldsymbol{S}(\omega)$ at three 
 different frequencies (see legends;
symbols stand for simulation results and lines for theoretical predictions). The inset in \textbf{D1} displays the effective coupling strength, \( g(\omega) \), at each frequency, highlighting that the ``most critical" dynamics occurs at \( \omega = 0 \). Power-law decay, signaling critical behavior, is observed at $0$ frequency in \textbf{D1} and for frequency $2\beta g$  in \textbf{D2}.
}
    \label{fig. paradigms}
\end{figure*}
 {{In the first example}}, $\boldsymbol{W}$ has independent i.i.d. random entries extracted from a zero-mean Gaussian distribution with variance $1/\sqrt{N}$ so that its eigenvalues obey the circle law for large $N$'s, i.e. they are uniformly distributed within a circle in the complex plane (Fig.\ref{fig. paradigms}A1) \cite{vM}. Representative time series and averaged power spectra are shown in  Fig.\ref{fig. paradigms}B1/C1.  The eigenvalue distribution of $\boldsymbol{S}(\omega=0)$ in Eq.(\ref{Sw}) has been explicitly computed \cite{HS} (see also SI). 
Its most salient features are shown in  Fig.\ref{fig. paradigms}D1: (i) 
the support of the distribution depends on $g$ and becomes unbounded from above, i.e. $\lambda_{max} \rightarrow \infty$ as $g\rightarrow 1$ and (ii) it decays asymptotically as a power law with exponent $-5/3$ 
as $g\rightarrow 1$, while there are strong $g$-dependent cutoffs for $g<1$. Inspection of Eq.(\ref{formula}) readily reveals that the same results apply to a generic $\boldsymbol{S}(\omega)$ just by replacing $g$ by an effective frequency-dependent coupling $g(\omega)=g/\sqrt{1+\omega^2}$ \cite{HS}. Thus, for the considered connectivity matrix following the circle law, where the first eigenvalues that become nearly unstable as \( g \rightarrow 1 \) are real-valued, the distribution of eigenvalues approaches criticality more closely at frequency \( \omega = 0 \) than at any other frequency (see inset of Fig. \ref{fig. paradigms}D1). However, this is not always the case, as we now illustrate.

{{In the second example}}, $\boldsymbol{W}$ is chosen as
\begin{equation}
\label{Oscillatory}
    \boldsymbol{W} =  \left [ \begin{array}{cc}
        \alpha \boldsymbol{J} & - \beta \boldsymbol{J}  \\
         \beta \boldsymbol{J} & \alpha \boldsymbol{J} 
    \end{array} \right ]
\end{equation}
where $\boldsymbol{J}$ is a symmetric random matrix with zero-mean and $1/\sqrt{N}$ variance  Gaussian entries. Its associated eigenvalues are real-valued and follow the semicircle law \cite{vM}, with the largest eigenvalue $\lambda_{J,max}=2$.  Note that, Eq.(\ref{Oscillatory}), even if inspired in networks with excitatory and inhibitory subpopulations, lacks specific significance as an actual neural connectivity model; it has been designed to ensure the presence of complex leading eigenvalues. Indeed, each real eigenvalue of
$\boldsymbol{J}$, $\lambda_J$ is mapped into two eigenvalues of
$\boldsymbol{W}$, $\lambda_W=(\alpha \pm i\beta) \lambda_J$, so that the eigenvalues with the largest real parts are $\lambda_{W,max} = 2 (\alpha \pm  \beta i)$ (Fig.\ref{fig. paradigms}A2). Thus, the edge of instability is approached as $g \rightarrow 1/2\alpha$ and corresponds to a \emph{Hopf bifurcation} signaling the onset of oscillations with frequency $\omega_C=\pm 2g \beta$ (as illustrated in Fig.\ref{fig. paradigms}B2/C2). Expressions for the eigenvalue distribution of $\boldsymbol{S}(\omega)$ for any $\omega$ can be analytically derived (see SI).
Remarkably, only at characteristic frequency $\omega_C=\pm 2\beta g$ (see Fig.\ref{fig. paradigms}D2) the resulting distribution exhibits a power-law tail (with exponent $-7/4$) as the edge of instability $2g-\frac{1}{\alpha}=0$ is approached. Therefore, the FDC matrix, $\boldsymbol{S}(\omega)$, allows us to observe fingerprints of criticality in data sets with a non-trivial temporal structure and unveil dynamical features that could not be uncovered through standard analyses of the integrated long-time covariance. 

\begin{figure*}[tbh!]
 \centering
    \includegraphics[width = 1.0\linewidth]{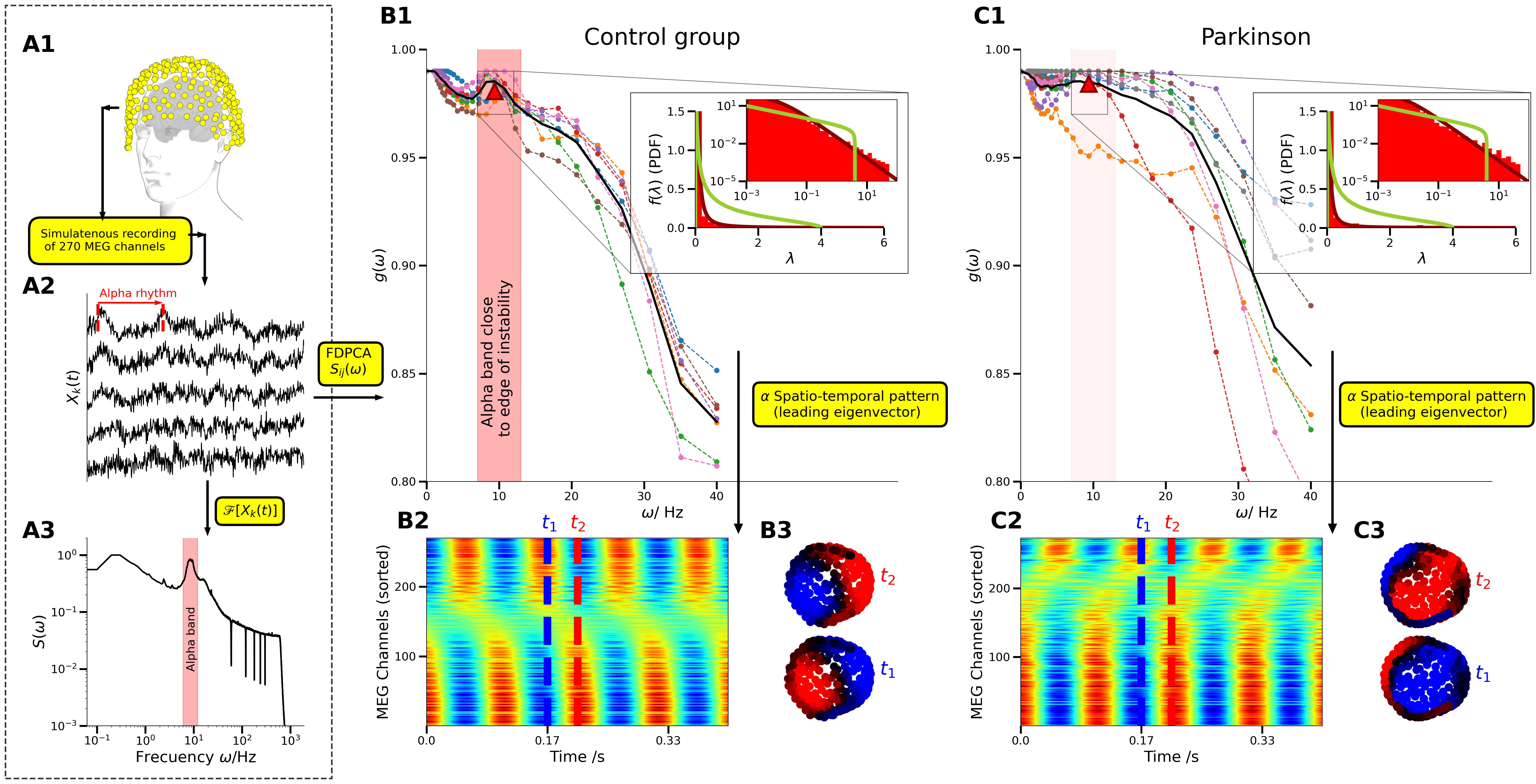}
    \caption{\textbf{Frequency-dependent covariance analyses reveal close-to-critical dynamics at different bands.}   \textbf{A1.} Illustration of a subject with the MEG helmet recording the activity of $270$ different channels distributed around the head. \textbf{A2.} Representative time series at $5$ arbitrary locations, exhibiting rich oscillatory behavior. \textbf{A3.} Typical power spectrum for one subject, showing a clear peak in the alpha band ($8-13$ Hz). \textbf{B. Analysis of the control group.} \textbf{B1.} Main plot: inferred value, $g(\omega)$, for different subjects (color-coded) and frequencies $\omega$: the dynamics are closest to criticality at very small frequencies and within the alpha band (the black curve stands for the group average). Inset: best fit of the empirical data to the theoretical distribution \cite{HS} for one representative subject (dark-red), together with the best fit to the Marchenko-Pastur distribution for residual correlations in random timeseries (green) both in linear and log-log scale. 
\textbf{B2.} Raster plot of the spatio-temporal pattern associated with the leading eigenvector of $\boldsymbol{S}(\omega)$ at the closest-to-critical frequency (alpha band). Channels have been sorted according to their relative phases. \textbf{B3.} Spatio-temporal pattern of \textbf{B2} represented at two different times $t_1$ and $t_2$, with the actual 3D locations of channels in the helmet (top view). Red (blue) colors stand for positive (negative) values. \textbf{C. Group with Parkinson's disease:} \textbf{C1.} As \textbf{B1}, but revealing a different organization of the distance to criticality across frequency bands. The key feature is that the peak in the alpha band in B1 is replaced by a less pronounced but much broader range of frequencies with close-to-critical behavior. \textbf{C2/C3.} As \textbf{B2/B3} but with distorted spatio-temporal patterns.}
    \label{fig. empirical}
\end{figure*}

Next, we demonstrate the potential of this framework by not only examining eigenvalues but also leveraging eigenvectors for the analysis of an extensive dataset of MEG recordings from both healthy controls and subjects with Parkinson's disease (see Fig.\ref{fig. empirical}A1/A2). For each subject, we computed the power-spectrum $S_{ii}(\omega)$ averaging over MEG recording channels.
Two frequency bands contribute the most to the spectral power: slow delta frequencies ($0.1-1$ Hz) and faster alpha waves ($8-13$ Hz), as shown in Fig.\ref{fig. empirical}A3. 
For each individual, we computed $\boldsymbol{S}(\omega)$ and the associated frequency-dependent eigenvalue distributions, which we then used to infer the effective coupling $g(\omega)$ at each frequency (see SI for the fitting procedure). Results for healthy (control) subjects are shown in Fig.\ref{fig. empirical}B1, revealing that closest-to-critical behavior ($g(\omega)\approx1$) is observed for very small frequencies (delta band) and in the alpha band. Moreover, as illustrated in the inset of Fig.\ref{fig. empirical}B1 for one individual, the eigenvalue distribution in the alpha band decays as a power-law (with the predicted exponent $-5/3$) and deviates from the expectation for uncorrelated random time series (i.e. from the Marchenko-Pastur distribution \cite{Tao}). Importantly, all these effects are lost if time series are randomized, even if their power spectra are preserved (see SI). Additionally, we determined the leading (complex) eigenvectors in each band to decompose the overall spatio-temporal activity patterns. In Fig.\ref{fig. empirical}B2, we illustrate this by showing the projection of the activity on the first complex eigenvector in the alpha band ($\sim 10$ Hz), revealing the presence of a travelling wave propagating along the occipital-frontal axis (see Fig.\ref{fig. empirical}B3 for a brain projection at two times separated by half a period). Similar analyses can be performed for other eigenvectors and frequencies, allowing for a full decomposition into a basis of spatio-temporal patterns. 

Remarkably, the study of subjects affected by Parkinson's disease reveals that there is a significantly broader interval of frequencies close to the edge of instability compared to the control group (see Fig.\ref{fig. empirical}C1). 
Although the corresponding eigenvalue distributions within the alpha band are well-fitted by power laws with the same exponent as observed in healthy patients (Fig.\ref{fig. empirical}C1 inset), the associated complex eigenvectors exhibit significant differences compared to those of the control group. As a consequence, the spatio-temporal waves show a significantly distinct structure in these patients: as illustrated in Fig.\ref{fig. empirical}C2 and Fig.\ref{fig. empirical}C3 the level of simultaneous activation (i.e. synchrony) between nodes within the dominant alpha wave is notably higher in Parkinson's patients compared to controls.
Although we do not delve deeper into this comparison, there is no doubt that decomposing brain activity into spatio-temporal principal components using a well-characterized basis presents promising avenues for future research in characterizing both healthy and pathological states.

In summary, we (i) present and leverage a tool (FD-PCA) to disentangle the sources of co-variability in high-dimensional complex timeseries, (ii) show how to estimate the effective dynamical regime at diverse frequency bands, allowing us to extend the concept of critical behavior to multi-frequency analyses, and (iii) devise a method to decompose spatio-temporal patterns of activity on a basis of frequency-dependent spatial waves, providing a dynamical generalization of  resting state networks. 
We are confident that this framework will have broad applications in neuroscience and beyond.

{\bf{Acknowledgments:}} 
We acknowledge the Spanish Ministry of Research and Innovation and
Agencia Estatal de Investigación (AEI),
MICIN/AEI/10.13039/501100011033, for financial support through Project
Ref. PID2020-113681GB-I00.


%

\onecolumngrid
\section*{Supplemental Information}
\setcounter{section}{0} 
\setcounter{equation}{0} 

\counterwithout{equation}{section}  

\renewcommand{\thesection}{S.\Roman{section}}
\renewcommand{\theequation}{S\arabic{equation}}
\renewcommand{\thetable}{S\arabic{table}}
\renewcommand{\thefigure}{S\arabic{figure}}

\onecolumngrid
\section{Covariance matrix and the Power Spectrum}
Given a collection of time series, $x_i(t)$, where $1\leq i\leq N$,  
of a stochastic process that is weakly stationary, the equal-time covariance matrix:
\begin{equation}
    C_{ij}=\langle x_i(t) x_j(t)\rangle,
\end{equation}
measures to what extent the behavior of the dynamical unit, $x_j(t)$, might be inferred from knowledge of $x_i(t)$ using simple linear regression. A typical technique of dimensionality reduction commonly used in data analysis involves projecting the time series activity, \( x_i(t) \), onto the eigenvectors (or principal components) of the matrix \( C_{ij} \) \cite{jolliffe_principal_2002}. These principal components represent the directions in which the dataset exhibits maximum variance.

In multidimensional complex systems, like the brain, information that propagates from one dynamical unit to another, carries a (small) delay, or time lag, \( \tau \), such that the behavior of \( x_j(t + \tau) \) can be inferred from the behavior of \( x_i(t) \) (see, for instance, \cite{Buzsaki2023}). In this case, the mathematical object measuring this linear dependence between timeseries is the time-lagged covariance matrix:
\begin{equation}
    C_{ij}(\tau)=\langle x_i(t)x_j(t+\tau) \rangle \ ,
\end{equation}
which, for weakly stationary stochastic processes, does not depend on $t$, but only on the time difference, $\tau$. This matrix is not generally symmetric (\( \langle x_i(t)x_j(t+\tau)\rangle \neq \langle x_j(t)x_i(t+\tau)\rangle \)), which means that eigenvalues and eigenvectors are both complex. This complexity makes them difficult to analyze.

We propose to use the Fourier transform of this time-lagged covariance matrix, also called the power-spectrum matrix or simply the frequency-dependent covariance (FDC) matrix, which ---by means of the Wiener-Kinchin theorem \cite[p. 17]{gardiner_handbook_2004}--- can be written as:
\begin{equation}
\label{WK theorem}
    S_{ij}(\omega)=\mathcal{F}[C_{ij}(\tau)](\omega)=\lim_{T\rightarrow\infty} \frac{1}{T}\langle X_{i,T}^*(\omega) X_{j,T}(\omega)\rangle  \ ,
\end{equation}
where $X_{i,T}(\omega)$ represents the Fourier coefficient of the time series, $x_i(t)$, at the particular frequency, $\omega$, in the time window $[0,T]$, i.e.,
\begin{equation}
    X_{i,T}(\omega)=\int_{0}^T e^{-i\omega t}x_i(t)dt \ ,
\end{equation}
and $(\circ)^*$ represents the complex conjugate of $(\circ)$. The FDC matrix has the advantage of having real eigenvalues since it is a Hermitian matrix. However, its eigenvectors are generally complex, as discussed in the main text, which leads to the emergence of spatio-temporal patterns. It's worth noting that the FDC matrix quantifies how much information from the Fourier transformed series \(X_{i,T}(\omega)\) can be used to reconstruct the Fourier transformed series \(X_{j,T}(\omega)\) through a straightforward linear regression.

\section{Frequency-dependent covariance matrix for the firing-rate model}
Let us consider an ensemble of $N$ neurons connected in a recurrent network with weights $W_{ij}$.  The firing-rate model serves as a parsimonious representation of the activity within these interconnected units, and is defined by a system of stochastic differential equations:
\begin{equation}
    \dot{x}_i(t)=-x_i(t)+ g \sum_{j=1}^N W_{ij} F\left(x_j(t)\right)+\sigma \xi_i(t) \; ,
    \label{FR}
\end{equation}
where $x_i(t)$ is the firing rate of neuron $i$, $F(x)$ is a gain or saturation function, $g$ is the so-called coupling strength ---which acts as an external parameter to control the phase at which the system is posed---, $\xi_i(t)$ is a Gaussian white noise with variance $\sigma^2$ (that will be set to $1$ for simplicity unless stated otherwise). The function $F(x)$ is typically assumed to be of the type $\tanh(x)$ \cite{HS}.  If $F(0)=0$, then the solution $x_i(t)=0$ is a fixed point of the deterministic dynamics, representing a state of quiescence or fluctuating background activity with $0$ mean. Linearizing around this fixed point and assuming the necessary conditions of differentiability of the gain function, allows us to approximate the non-linear model by a simpler model of linearly-coupled units (the Ornstein-Uhlenbeck process):
\begin{equation}
\label{LFR}
    \dot{x}_i(t)=-x_i(t)+ \overbrace{F'(0)}^{=1}g \sum_{j=1}^N W_{ij} x_j(t)+\sigma^2\xi_i(t)\; .
\end{equation}
The Ornstein-Uhlenbeck process serves as an approximation for the behavior of the non-linear system defined by equation (\ref{FR}) in the quiescent phase. Indeed, let $\lambda_{max}$ denote the eigenvalue with largest real part of the matrix $\mathbf{W}$. Then, the condition of linear stability demands that $g\lambda_{max}<1$. The point $g=1/\lambda_{max}$ marks the so-called ``edge of instability", such that, beyond this point, the dynamics are no longer properly described by the linearized system.

The stationary solution of the linear model, Eq.(\ref{LFR}), can be computed as:
\begin{equation}
    \mathbf{x}(t)=e^{-\mathbf{A}t}\int_{-\infty}^t e^{\mathbf{A}s}\boldsymbol{\xi}(s)ds \ ,
\end{equation}
where $\mathbf{A}=-(\mathbf{I}-g\mathbf{W})$ and $\mathbf{I}$ is the identity matrix. Using this expression, it is straightforward to prove that the equal-time covariance matrix, $\mathbf{C}=\langle \mathbf{x}(t)\mathbf{x}^T(t)\rangle $, verifies a Lyapunov equation \cite{Barnett}:
\begin{equation}\mathbf{A}\mathbf{C}+\mathbf{C}\mathbf{A}^T=\int_{-\infty}^t \mathbf{A} e^{-\mathbf{A}(t-s)}e^{-\mathbf{A}^T(t-s)}ds+\int_{-\infty}^t e^{-\mathbf{A}(t-s)}e^{-\mathbf{A}^T(t-s)}\mathbf{A}^T ds=\int_{-\infty}^t \frac{d}{ds}\left[ e^{-\mathbf{A}(t-s)}e^{-\mathbf{A^T}(t-s)} \right]=\mathbf{I} \ .
\end{equation}
On the other hand, the time-lagged covariance matrix can be computed as:
\begin{equation}
    \mathbf{C}(\tau)=\langle \mathbf{x}(t)\mathbf{x}(t+\tau)\rangle = \int_{-\infty}^{t} e^{-\mathbf{A}(t-s)}e^{-\mathbf{A}^T(t+\tau-s)} ds = \begin{cases}
      \mathbf{C}e^{-\mathbf{A}^T\tau}, &\text{   if  } \tau>0 \\
      e^{\mathbf{A}\tau} \mathbf{C}, &\text{   if  } \tau<0      
    \end{cases} 
\end{equation}
Hence, the frequency-dependent covariance matrix can be exxpressed as
\begin{equation}
    \mathbf{S}(\omega)=\int_{-\infty}^{\infty} \mathbf{C}(\tau)e^{-i\omega \tau} d\tau=\int_0^{\infty} \mathbf{C}e^{-\mathbf{A}^T\tau} e^{-i\omega\tau}d\tau+\int_{-\infty}^0 e^{\mathbf{A}\tau} \mathbf{C} e^{-i\omega\tau}d\tau=\frac{1}{\mathbf{A}+i\omega\mathbf{I}}\mathbf{C}+\mathbf{C}\frac{1}{\mathbf{A}^T-i\omega\mathbf{I}} \ .
\end{equation}
Finally, using the Lyapunov equation, we get to:
\begin{equation}
    (\mathbf{A}+i\omega\mathbf{I})\mathbf{S}(\omega)(\mathbf{A}^T-i\omega\mathbf{I})=\mathbf{A}\mathbf{C}+\mathbf{C}\mathbf{A}^T=\mathbf{I} \ ,
\end{equation}
so that
\begin{equation}
\label{S}
    \mathbf{S}(\omega)=\frac{1}{\mathbf{A}+i\omega\mathbf{I}} \times \frac{1}{\mathbf{A}^T-i\omega\mathbf{I}} \ .
\end{equation}

\section{Power-law behavior in the eigenvalues of the FDC matrix}
Consider again the linear firing-rate model defined by Eq.(\ref{LFR}). In \cite{HS}, the authors derived an analytical expression for the distribution of eigenvalues of the so-called ``long-time window covariance matrix" (i.e., the FDC matrix at frequency $\omega = 0$) when the weights, $W_{ij}$, are independently drawn at random from a normal distribution with $0$ mean and variance $1/\sqrt{N}$:
\begin{equation} \label{Eq. H&S}
       f(x)=  \frac{3^{\frac{1}{6}}}{2\pi g^2 x^2} \left[ \sum_{\xi=1,-1} \xi\left( \left(1+\frac{g^2}{2}\right)x-\frac{1}{9} +\right.\right.       \left.\left.\xi\sqrt{\frac{(1-g^2)^3 x(x_+-x)(x-x_-)}{3}}    \right)^{\frac{1}{3}} \right] \ .
\end{equation}
where $ x\in [x_-, x_+]$ and the support limits are defined as:
\begin{equation}
x_{\pm}=\frac{2+5g^2-g^4/4\pm \frac{1}{4}g(8+g^2)^{3/2}}{2(1-g^2)^3} \ .
\end{equation}
Furthermore, it can also be easily proved from Eq.(\ref{S}) that the FDC matrix, $\mathbf{S}(\omega)$, has a similar behavior for $\omega\neq 0$, with the coupling strength $g$ replaced by an effective coupling, $g(\omega)$, given by:
\begin{equation}
    g(\omega)=\frac{g}{\sqrt{1+\omega^2}} \ ,
\end{equation}
which implies that critical behavior is always observed first at frequency $\omega=0$. Inspired by this idea, in this paper we built a synaptic matrix $\mathbf{W}$ that induces oscillatory behavior, such that the first eigenvalues to become unstable have nonzero imaginary part. In this way, the spectrum of eigenvalues of the FDC matrix becomes a power law (i.e. shows the fingerprints of criticality) only at the particular frequency of oscillation. To pursue this idea, we looked for a Wilson-Cowan-like model with two populations of neurons, $\mathbf{E}(t)$ and $\mathbf{I}(t)$, each of them consisting of $N$ units. The most general expression for a model of this type ---assuming that the refractory period is $0$--- would be:
\begin{equation}
    \begin{cases}
        \dot{E}_k(t)=-E_k(t)+g\alpha_{EE}\sum_{j=1}^N J^{EE}_{kj} S(E_j(t))+g\alpha_{EI}\sum_{j=1}^N J^{EI}_{kj} S(I_j(t))+\xi_k(t) \\
        \\
        \dot{I}_k(t)=-I_k(t)+g\alpha_{IE}\sum_{j=1}^N J^{IE}_{kj} S(E_j(t))+g\alpha_{II}\sum_{j=1}^N J^{II}_{kj} S(I_j(t))+\eta_k(t) \ ,
    \end{cases}
    \label{oscillatory matrix SI}
\end{equation}
where the scalars $\alpha_{AB}$, $A,B\in \{E,I\}$, denote the coupling strengths of the (directed) connections from population $B$ to population $A$, and $\mathbf{J}^{AB}$ stands for the synaptic weight matrix mediating the interaction of each pair of neurons from population $B$ towards $A$. For simplicity, and in order to make the calculations feasible, we chose all the synaptic weight matrices to be symmetric and identical, $\mathbf{J}^{AB}=\mathbf{J}$, with random entries drawn from a Gaussian distribution with $0$ mean and variance $1/\sqrt{N}$. On the other hand, we chose $\alpha^{EE}=\alpha^{II}=\alpha$ and $\alpha^{IE}=\beta=-\alpha^{EI}$, leading to the matrix shown in the main text (see Eq.(\ref{Oscillatory})). This choice of parameters will make calculations particularly simple. However, notice that this model does not represent excitatory and inhibitory units because the inhibitory weights ($\alpha^{IE}$ and $\alpha^{II}$) should always have negative signs, while excitatory weights ($\alpha^{EE}$ and $\alpha^{EI}$) should be positive. In any case, it is worth noting that the qualitative behavior of this family of models, regardless of the choice of the synaptic couplings, is very similar in the sense that they all exhibit power-law behavior in the distribution of eigenvalues of the frequency-dependent covariance matrix only at the characteristic frequency at which the corresponding Hopf-bifurcation happens. 
Let us then focus on the matrix with the shape proposed in the main text:
\begin{equation} \label{OscillatorySI}
    \boldsymbol{W} =  \left [ \begin{array}{cc}
\alpha \boldsymbol{J} & - \beta \boldsymbol{J}  \\
         \beta \boldsymbol{J} & \alpha \boldsymbol{J} 
    \end{array} \right ] \ .
\end{equation}
Observe that this case is simpler to solve because $\mathbf{W}$ and $\mathbf{W}^T$ commute, meaning that they are both simultaneously diagonalizable, thus making it easier to diagonalize $\mathbf{S}(\omega)$, which involves terms of the type $\mathbf{W}\mathbf{W}^T$. Indeed, let $\mathbf{w}\in \mathbb{C}^{2N}$ be a vector of the type
\begin{equation}
    \mathbf{w}=\left [ \begin{array}{c}
        \mathbf{v}   \\
        k \mathbf{v}\,
    \end{array} \right ]
\end{equation}
where $\mathbf{v}\in\mathbb{R}^N$ is an eigenvector of the matrix $\mathbf{J}$ with eigenvalue $\mu$ associated and $k\in\mathbb{C}$ is a scalar. We force $\mathbf{w}$ to be an eigenvector of the matrix $(\mathbf{A}+i\omega)$, with associated eigenvalue $\lambda_1$, leading to a system of equations for both $k$ and  $\lambda_1$:
\begin{equation}
    (\mathbf{A}+i\omega)\mathbf{w}=\lambda\mathbf{w}\Rightarrow 
    \begin{cases}
        1+i\omega-\alpha\mu+\beta k\mu&=\lambda_1 \\
        -\beta\mu+k(1+i\omega-\alpha\mu)&=k\lambda_1 .
    \end{cases}
    \label{system of eqs}
\end{equation}
This system of equations leads to $k=\pm i$ and $\lambda_1=(1-\alpha\mu)+(\omega\pm \beta\mu)i$. On the other hand, since $\mathbf{W}$ and $\mathbf{W}^T$ are simultaneously diagonalizable, so are $(\mathbf{A}+i\omega)$ and $(\mathbf{A}^T+i\omega)$, which means that the eigenvector $\mathbf{w}$ is also an eigenvector of $(\mathbf{A}^T-i\omega)$. Indeed, for this matrix, imposing that $\mathbf{w}$ is an eigenvector of $(\mathbf{A}^T-i\omega)$ leads to a system of equations similar to Eq.(\ref{system of eqs}) with $\omega \rightarrow -\omega$ and $\beta\rightarrow -\beta$. Hence, we deduce that $\mathbf{w}$ is an eigenvector with eigenvalue $\lambda_2=(1-\alpha\mu)-(\omega\pm \beta\mu)i$. Since we are interested in the eigenvalues of the FDC matrix, we finally conclude that $\mathbf{w}$ must be an eigenvector of $\mathbf{S}(\omega)$ with the following eigenvalue $\lambda_\pm$:
\begin{equation}
    \lambda_{\pm}=\frac{1}{(1-\alpha\mu)^2+(\omega\pm\beta\mu)^2} \ .
    \label{transformation}
\end{equation}
Thus, we have an explicit transformation from the set of $N$ eigenvalues, $\mu$, of the matrix $\mathbf{J}$ to the set of $2N$ eigenvalues, $\lambda_\pm$, of the matrix $\mathbf{S}(\omega)$. Observe that the mapping has two branches, a positive one and a negative one, leading to a doubling of the number of eigenvalues. To get the density of the eigenvalues $\lambda_\pm$ it suffices to apply a change of variables to the eigenvalues $\mu$, which are distributed according to the semi-circle law \cite{HS} (due to the symmetry of the matrix $\mathbf{J}$). In our case, the transformation defined by Eq.(\ref{transformation}) is non-monotonic, so one needs to use the theorem of change of variables in every subset where the transformation becomes a bijection. Inverting this transformation yields:
\begin{equation}
\label{inverse transformation}
    \mu^\xi_\eta(\lambda)=\frac{1}{\alpha^2+\beta^2}\left[ (\alpha-\xi\omega\beta)+\eta\sqrt{\frac{1}{\lambda}(\alpha^2+\beta^2)-(\beta+\xi\omega\alpha)^2}\right] \ ,
\end{equation}
with $\xi,\eta\in\{-1,+1\}$. The Jacobian of this local inverse is:
\begin{equation}
    |J^\xi(\lambda)|=\frac{1}{\lambda^2\sqrt{\frac{\alpha^2+\beta^2}{\lambda}-(\alpha\omega+\xi\beta)^2}}
\end{equation}
so that the final density of eigenvalues $\lambda$ of the FDC matrix $\mathbf{S}(\omega)$ can be expressed as:
\begin{equation}
    \rho(\lambda)=\frac{k}{2}\left[ \sum_{\xi,\eta=\pm} \left( \sqrt{4g^2-\mu_\eta^\xi(\lambda)^2} \right)|J^\xi(\lambda)| \right]
    \label{density lambda} \ ,
\end{equation}
where $k=1/2\pi g^2$ and the factor $1/2$ arises as a normalization factor when summing the contribution of all the local inverses (sum over the two branches $\xi=\pm 1$). The local inverses, on the other hand, are defined in the following set:
\begin{equation}
    D^\xi=\left\{\lambda\in\mathbb{R}: \Delta^\xi=\frac{1}{\lambda}(\alpha^2+\beta^2)-(\beta+\xi\omega\alpha)^2>0\right\},
\end{equation}
a condition imposed on the discriminant $\Delta^\xi$ of the inverse transformation of Eq.(\ref{inverse transformation}). This leads to a maximum eigenvalue given by:
\begin{equation}
    \lambda<\lambda^\xi_{max,1}=\frac{(\alpha^2+\beta^2)}{(\beta+\xi\omega\alpha)^2} \ .
\end{equation}
Alternatively, the expression in Eq.(\ref{density lambda}) is only defined whenever $4g^2-\mu_\eta^\xi(\lambda)^2>0$, which implies the following bound:
\begin{equation}
    \lambda<\lambda_{max,2}^\xi=\frac{(\alpha^2+\beta^2)}{(\beta+\xi\omega\alpha)^2+(\alpha-\xi\omega\beta)-2g(\alpha^2+\beta^2)} \ .
\end{equation}
Observe that, in order to have a diverging support (a necessary condition to observe a power-law), one must have that both of these bounds go to infinity, i.e., $\lambda^\xi_{max,j}\rightarrow\infty$ ($j=1,2$), which only occurs if $\beta+\xi\omega\alpha=0$ (in order to ensure that $\lambda_{max,1}\rightarrow \infty$) and $1/\alpha=2g$ (in order to ensure that $\lambda_{max,2}\rightarrow \infty$). The first condition proves that there are only two characteristic frequencies (positive and negative) where one can observe diverging eigenvalues:
\begin{equation}
    \omega_C=\pm\frac{\beta}{\alpha} \ ,
\end{equation}
which are precisely the imaginary part of the leading eigenvalues (complex conjugates of one another) of the coupling matrix $\mathbf{W}$. The second condition, on the other hand, ensures that the real part of the leading eigenvalue of the matrix $\mathbf{W}$ is at the edge of instability. Hence, at the characteristic frequency, $\omega_C$, we observe oscillatory behavior arising from a Hopf bifurcation and the FDC matrix has a density of eigenvalues that becomes unbounded from above. The preceding arguments prove that the necessary conditions for the existence of a power law are precisely $\beta+\xi\omega\alpha=0$ and $1/\alpha=2g$. One of these conditions determines the characteristic frequency of oscillations, while the other sets the edge of instability. 

To prove that the eigenvalue distribution of the FDC matrix is a power-law we plug this conditions into the density function described by Eq.(\ref{density lambda}) and take the limit $\lambda\rightarrow\infty$, which leads to:
\begin{equation}
    \rho(\lambda)\sim \frac{1}{\lambda^2\lambda^{-1/2}}\sqrt{\left(\frac{-2\eta^2}{\lambda(\alpha^2+\beta^2)}-\frac{2\eta}{\alpha\sqrt{\lambda(\alpha^2+\beta^2)}}\right)}\sim\lambda^{-7/4}.
\end{equation}
This shows that the density of eigenvalues is scale-invariant with a power exponent of $-7/4$, which coincides with the exponent of the eigenvalue distribution of the long-time window covariance (namely, the power spectrum at frequency $\omega=0$) for the symmetric Gaussian ensemble \cite{HS}.

\section{Effects of non-linearities}
\begin{figure*}[tbh!]
\centering
    \includegraphics[width = 1.0\linewidth]{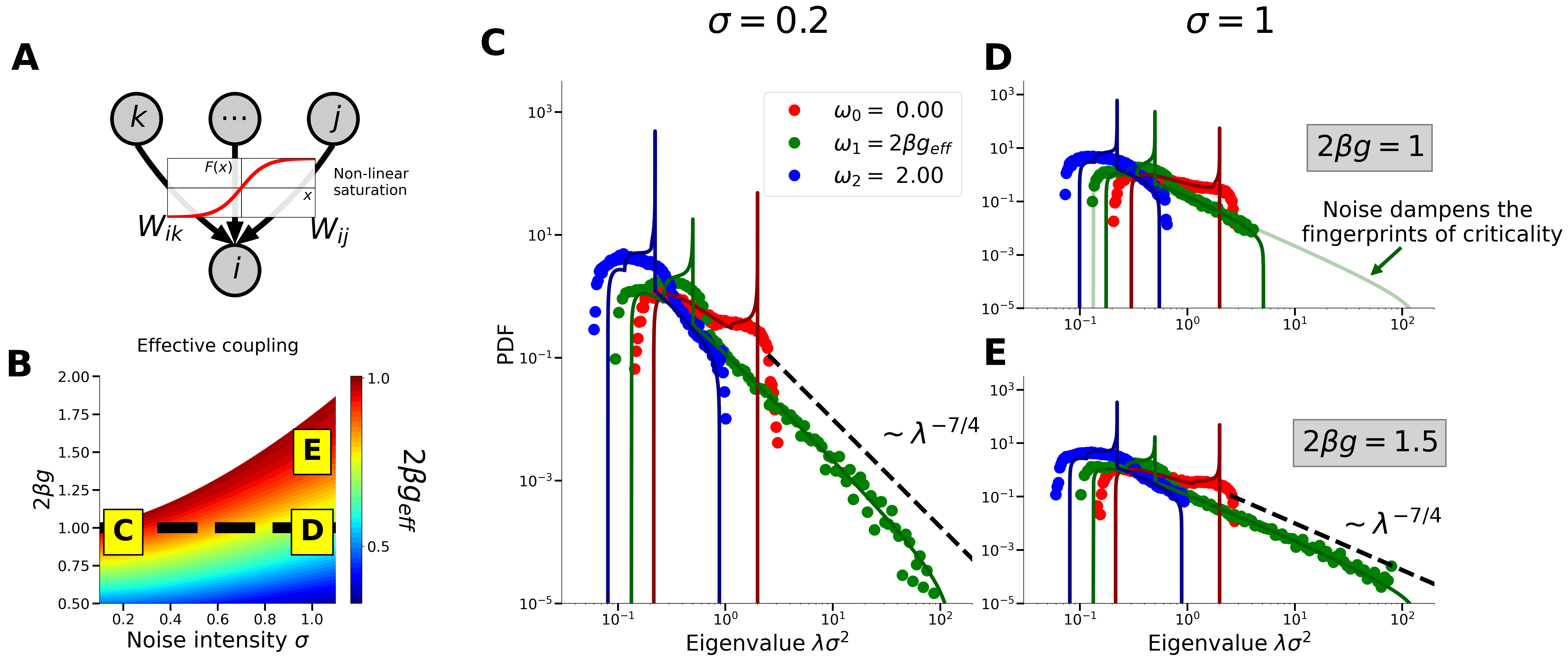}
    \caption{\textbf{Non-linearities effectively shift the coupling strength at which critical behavior is observed.}   \textbf{A}. Scheme of the recurrent neural network of firing-rate units with non-linear saturation given by $F(x)=\tanh(x)$: the input to each neuron $k$ is properly weighted by the synaptic strength $W_{ik}$ and transformed through the gain function $F(x)$ before arriving at neuron $i$. \textbf{B}. The effective coupling strength as a function of $g$ and the noise intensity $\sigma$, calculated using the \textit{ansatz} $g_{eff}=g\langle F'(x_i)\rangle$. As we can see, the larger the noise intensity, the larger the coupling strength $g$ needed to observe an effective critical coupling $g_{eff}$ near to the inset of criticality. The squares mark the values of $(2\beta g,\sigma)$ for each of the subplots: $(2\beta g,\sigma)=(1,0.2)$ (subplot C), $(2\beta g,\sigma)=(1,1)$ (subplot D) and  $(2\beta g,\sigma)=(1.5,1)$ (subplot E). The black dotted line marks the onset of instability in the linear theory, occurring when the eigenvalue with the largest real part (in our case, this real part is precisely $2\beta g$) is equal to $1$. \textbf{C}. Eigenvalue density of the frequency-dependent covariance matrix at three different frequencies, showing power-law behavior at the characteristic frequency $\omega_1$. Dots are the results from simulations, while lines correspond to theory. \textbf{D}. Same plot as in C, but for a noise intensity of $\sigma=1$, showing how, at the characteristic frequency, the fingerprints of criticality have vanished as noise effectively ``dampens" them \cite{Helias-noise}. \textbf{E}. Same plot as in D, but for a value of $2\beta g=1.5$ (beyond the edge of linear instability). For this value of the coupling strength, we recover a power-law in the distribution of eigenvalues of the frequency-dependent covariance matrix with the same decaying exponent, since $2\beta g_{eff}$ gets close to $1$.}
    \label{fig. nonlinear}
\end{figure*}

Up to this point, we have solely examined signatures of criticality in the frequency-dependent covariance matrix for the linear model, where the gain function in Eq.(\ref{FR}) is defined as the identity function. 
Here, we delve into the scenario where $F(x)=\tanh(x)$, a commonly employed saturation function \cite{Crisanti88} (see Fig.\ref{fig. nonlinear}A). In the realm of nonlinear dynamics, analytical calculations remain confined to the average behavior of a typical neuron in the thermodynamic limit, often disregarding correlations \cite{Crisanti88,Martorell}. Hence, we have to resort to numerical examples. Following the approach of \cite{HS} and \cite{clark_dimension_2023}, we assert that the effective coupling strength in the non-linear model, denoted as $g_{\text{eff}}$, is linked to the average slope of the non-linearity function $F(x)$ over the firing rates $x_i(t)$ as: 
\begin{equation}
    g_{eff}=g\langle F'(x_i(t))\rangle \ ,
\end{equation}
where averages are to be taken over time, nodes, and structural disorder. Using this expression, we were able to compute numerically the value of $g_{eff}$ as a function of the coupling strength $g$ and the variance of the noise, $\sigma^2$, defined in Eq.(\ref{FR}) (see Fig.\ref{fig. nonlinear}B) in the case in which the synaptic weight matrix is set according to Eq.(\ref{oscillatory matrix SI}). As we can see, as one increases the noise intensity, $\sigma$, for a fixed value of $g$, there is a smaller effective coupling. For two different values of sigma, $\sigma=0.2$ and $\sigma=1$, we
repeated the same measurements as in Fig.\ref{fig. paradigms}D2 (see main text) to obtain the eigenvalue distribution of the frequency-dependent covariance matrix at different frequencies (see Fig.\ref{fig. nonlinear}C/D/E). We observed that, for $\sigma=1$, even when the system is posed at the edge of linear instability  ---in particular, for the simulations shown in subplots C/D, the largest eigenvalue of the matrix $\boldsymbol{W}$ has real part equal to $2\beta g=1$--- there are no fingerprints of criticality, i.e. the system is away from criticality, as noise effectively ``dampens" this power-law tail. In other words, the combined presence of non-linearities and noise, shifts the transition point --the edge of chaotic behavior-- to larger effective values of $g_{eff}$.

Instead, to recover the scale-free behavior of the eigenvalues of the frequency-dependent covariance matrix, one needs to go beyond the limit of linear instability, i.e., to the region in which $2\beta g>1$. In particular, in Fig.\ref{fig. nonlinear}E we observe how, when the coupling strength $g$ is increased to $2\beta g=1.5$, one recovers the same scaling behavior and power-laws predicted by the linear theory.

\section{Data acquisition and processing}
\subsection{The OMEGA dataset}
The full OMEGA dataset consists of nearly 900 resting-state magnetoencephalography (MEG) recording sessions, for a total of over 75 hours of data \cite{niso_omega_2016}. The recordings were taken with a 275-channel 2005 series CTF MEG system at the McConnell Brain Imaging Center, at a time resolution of 2400Hz. The data is structured according to the BIDS 1.7.0 standard and was pre-processed with the open-source software Brainstorm \cite{tadel_brainstorm_2011}. The access to the dataset was granted for this project for 12 months, and reviewed and approved by the internal research ethics board.
Out of the total of 294 volunteering participants in the dataset, we selected 7 that were healthy controls and 8 that were diagnosed with Parkinson's disease. These were the subjects for which the timeseries were longest (in particular, they had a duration of 1200s), allowing a better estimate of the FDC matrix. 

Following the recommended protocol in \cite{tadel_brainstorm_2011}, we applied a Notch filter at frequencies of 60, 120, 180, 240 and 300Hz to remove the
noise due to the AC power line frequency in Canada, and a high-pass filter at 0.3Hz, 60dB. Moreover, since the dataset contains simultaneous bipolar Electrocardiogram (ECG) and vertical and horizontal bipolar Electro-oculogram (EOG) recordings, we cleaned the MEG data from the artifacts due to heartbeats and eye blinks. The artifact cleaning procedure consists of detecting reproducible stereotyped and localized topographies, that correlate with the signals of the EOG and ECG, through a Signal-Space Projection \cite{tadel_brainstorm_2011}.

\subsection{Measuring Frequency-Dependent Covariances}
\begin{figure*}[tbh!]
        \centering
    \includegraphics[width = 1.0\linewidth]{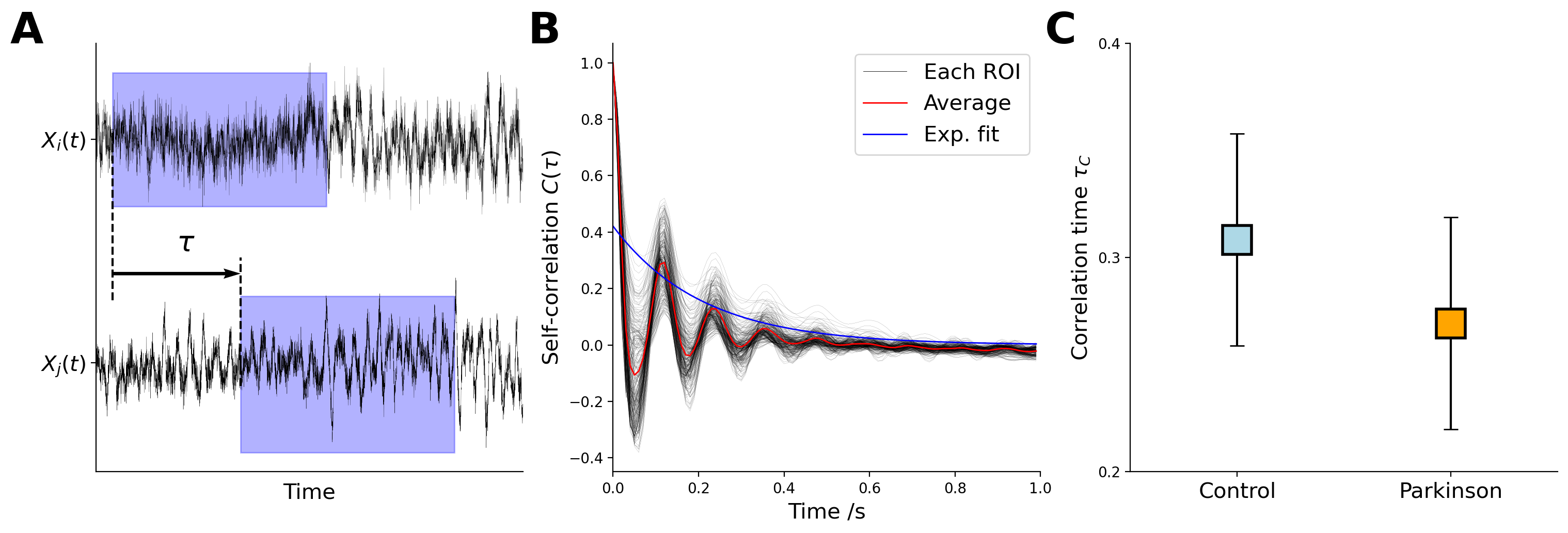}
    \caption{\textbf{Correlation time of the time series for both the control and Parkinson's group.}   \textbf{A}. Scheme of the procedure used to calculate the self-correlation function using a sliding window. The time-lagged correlation is $C_{ij}(\tau)=\langle x_i(t)x_j(t+\tau)\rangle$, where the average is performed over all the times $t\in[0,\Delta T]$, with $\Delta T$ denoting the width of the sliding window. \textbf{B}. Plot of the self-correlation function $C_{ii}(\tau)$ at each of the ROIs of the helmet, together with their group average (in red) for a randomly chosen patient of the control group. We also show the result of exponentially fitting this average self-correlation in order to get the correlation time. \textbf{C}. Correlation time $\tau_C$ for both the control group and the Parkinson's disease group. Error bars represent the standard deviation across patients.}
    \label{fig. corr time}
\end{figure*}

The frequency-dependent covariance matrix can be estimated from a sample of time series $x_i(t)$ ($1\leq i\leq N$) using two different procedures. The first one consists of measuring the time-lagged covariance between each pair of time series and Fourier transforming it. The problem with this method is that, for a finite sampling of the time series, the result will not, in general, be a Hermitian, positive-definite matrix as it should be. Only in the limit of infinite samples, the covariance matrix measured using this method will converge to a matrix displaying the correct properties. The second method, on the other hand, relies on the Wiener-Khinchin theorem to write the frequency-dependent covariance matrix $\mathbf{S}(\omega)$ as:
\begin{equation}
\label{Sw sup}
    S_{ij}(\omega)=\lim_{\Delta t\rightarrow\infty} \frac{1}{\Delta t}\langle X_{i,\Delta t}^*(\omega) X_{j,\Delta t}(\omega)\rangle  \ ,
\end{equation}
where:
\begin{equation}
    X_{i,\Delta t}(\omega)=\int_0^{\Delta t} x_i(t')e^{-i\omega t'}dt' \ .
\end{equation}
From Eq.(\ref{Sw sup}) it is clear that $\boldsymbol{S}(\omega)$ is a Hermitian matrix. To estimate $X_{i,\Delta t}(\omega)$ out of a time series $x_i(t)$ sampled a number $T$ of times (meaning that each data point can be written as $x_i(t_k)$, where $0\leq k \leq T-1$), what we do is to split the whole series in a total of $M$ ``chunks" of length $L$ (so that $T=L\cdot M$). For the $s^{th}$ chunk ($0\leq s\leq M-1$), which consists of a sample of the time series $x_i(t_k)$ for $s\cdot L\leq k\leq (s+1)\cdot L$, we can estimate $X_{i,\Delta t=L}(\omega)$ by means of the discrete Fourier transform:
\begin{equation}
    X_{i,s}(\omega)=\frac{1}{L}\sum_{k=s\cdot L}^{(s+1)\cdot L} x_i(t_k) e^{i\omega t_k}.
\end{equation}
If we denote as $\boldsymbol{X}(\omega)$ the $N\times M$ matrix whose elements correspond to the Fourier transform of the $i^{th}$ time series in the $s^{th}$ chunk, we can abbreviate the FDC matrix as:
\begin{equation}
    \boldsymbol{S}(\omega)=\frac{1}{M}\boldsymbol{X}(\omega)\boldsymbol{X}(\omega)^H.
\end{equation}
Writing it this way, it is also clear that the FDC matrix that this procedure estimates is positive-definite. Determining the optimal way to split the time series (i.e., to choose the length of the chunks) typically involves taking into account several factors:
\begin{enumerate}
    \item The samples-to-units ratio, $N/M$, which determines the number of non-zero eigenvalues (if $M>N$, the rank of the matrix is, at most, $N$, meaning there will be, at most, $N$ non-trivial eigenvalues) should be as big as possible.
    \item The duration of each chunk, $L$, is related to the smallest frequency that this method can resolve. If we denote as $\omega_0$ the smallest frequency that can be resolved using a DFT, one then finds that:
    \begin{equation}
        \omega_0=\frac{1}{2}\frac{f_S}{L} \ ,
    \end{equation}
where $f_S=1/(t_{k+1}-t_k)$ is the sampling frequency. Thus, $L$ has to be big enough so that the smallest frequencies can also be analyzed.
    \item The duration of the chunks, $\Delta t_C=L/f_S$, should be bigger than the correlation time of the time series in order to reduce correlations among the measurements $X_{i,s}$ and $X_{i,s'}$. It is simple to prove that, if the correlation of the time series decays exponentially with a correlation time $\tau_C$, then the correlation between chunks decays exponentially with a characteristic rate equal to $\Delta t_C/\tau_C$, so we choose $\Delta t_C\geq \tau_C$.
\end{enumerate}
In the case of the OMEGA dataset, the sampling frequency was $f_S=2400$ Hz, and we chose subjects with total recording time  of $1200$ s (merging different recording sessions). We studied the correlation time for different patients and different regions (see Fig.\ref{fig. corr time}) and we systematically observed that the latter was below $1$ s for both the control and Parkinson's groups. Since the number, $N$, of nodes was typically $270$ channels, we split the time series in chunks with a duration of $2.5$ s (larger than the correlation time), giving us a total of $M=480$ chunks (larger than the number of nodes).

\subsection{Fitting data to theory}
After estimating the FDC matrix, we calculated the effective coupling strength $g(\omega)$ by fitting the empirical distribution that we obtained to the analytical expression in Eq.(\ref{Eq. H&S})\cite{HS}. More specifically, we looked for the value of $g$ that minimized the quadratic error in the $L^2$ norm using the Cramer-von Mises statistics between the empirical cumulative distribution $F_n(x)$ and the theoretical one, $F(x)=\int_{-\infty}^x f(x)dx$ (where $f(x)$ is defined as in Eq.(\ref{Eq. H&S}) but with the correction for the finite sample-to-units ratio $\alpha=N/M$ derived in \cite{HS}):
\begin{equation}
    D^2_{CvM}=\int (F(x)-F_n(x))^2dF_n(x)=\frac{1}{12n^2}+\frac{1}{n}\sum_{i=1}^n\left(F(x_i)-\frac{2i-1}{2n}\right) \ ,
\end{equation}
where $n$ is the total number of samples and $x_i$ are the eigenvalues of the empirical frequency-dependent covariance matrix. We repeated this analysis for each subject and then calculated the average across subjects in each group.

\subsection{Criticality analysis on surrogated data}
\begin{figure*}[tbh!]
    \centering
    \includegraphics[width = 1.0\linewidth]{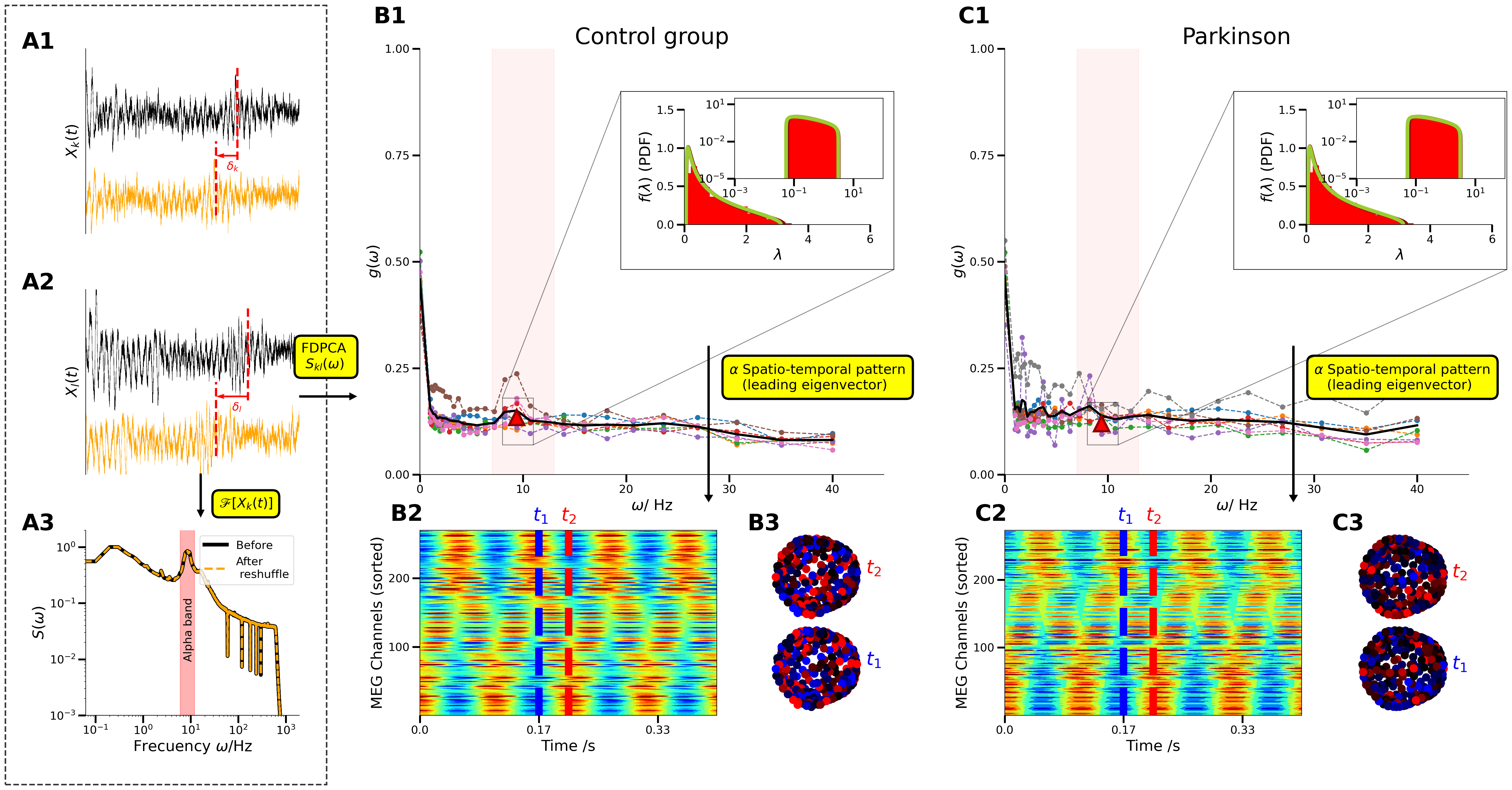}
    \caption{\textbf{Signatures of criticality in the frequency-dependent covariance analyses disappear when the time series are reshuffled.}   \textbf{A1/A2.} Illustration of the reshuffling procedure: each time series is shifted by a random amount $X_k(t)\rightarrow X_k(t+\delta_k)$. \textbf{A3.} This transformation preserves the self-correlation of each node, meaning that the power spectrum remains untouched. \textbf{B1/C1.} Distance to the edge of instability $g(\omega)$ for the control group (B1) and the Parkinson's group (C1) showing how the reshuffling leads to an effective coupling strength which is approximately $0$. The inset shows the distribution of eigenvalues for a value of $\omega\approx10 Hz$ (inside the alpha band) showing that the distribution is well captured by the Marchenko-Pastur distribution for uncorrelated variables (green curve). \textbf{B2/C2}. Raster plot of the spatio-temporal pattern associated with the leading eigenvector of the frequency-dependent covariance matrix $\boldsymbol{S}(\omega)$ at the alpha band, where channels have been sorted according to their phase difference. The pattern shows that, at each time, there are almost no nodes with the same phase, reflecting the fact that this pattern is pure noise. \textbf{B3/C3}. The spatio-temporal pattern of (B2/C2) projected over the actual 3d positions of the channels (top view). Red colours indicate positive values and blue colours indicate negative ones. The pattern shows no spatio-temporal structure whatsoever. }
    \label{fig. empirical_2}
\end{figure*}

As a sanity check, we performed the same analysis to measure the distance to criticality (i.e., the coupling strength $g(\omega)$) over surrogated data. To show that there is, in principle, no relation between $g(\omega)$ and the power spectrum, we shuffled randomly each of the time series such that the temporal structure (i.e., the self-correlation function $C_{ii}(\tau)=\langle X_i(t)X_i(t+\tau)\rangle$) was preserved. This is achieved, for instance, by shifting the time series at each node by a random amount: $X_k(t)\rightarrow X_k(t+\delta_k)$ (where $\delta_k$ are random shifts, see Fig.\ref{fig. empirical_2}A1/A2). This transformation granted the conservation of the power spectrum (see Fig.\ref{fig. empirical_2}A3), which is calculated as the Fourier transform of the self-correlation function for each node, averaged across nodes. However, because the reshuffling is carried out independently in each of the time series, we expect the frequency-dependent covariance matrix to reflect only noisy behavior. Indeed, the distance to the edge of instability $g(\omega)$ for the randomized data clearly shows that the timeseries become uncorrelated (see Fig.\ref{fig. empirical_2}B1/C1), while the leading spatio-temporal waves in the alpha band do not reveal any patterns of synchronized behavior (see Fig.\ref{fig. empirical_2}B2/B3/C2/C3). The only frequency at which we still observe some non-trivial correlations is at the long-time window ($\omega=0$, where $g(\omega)\approx 0.5$). This happens because, at frequency $0$, we are just integrating the time series over the whole chunk, so that there might remain some structure in the correlations even when the time series have been shifted. 

\end{document}